\documentclass[aps,prl,twocolumn,superscriptaddress]{revtex4}

\usepackage{amsmath,amsfonts,amssymb}
\usepackage{graphicx,float,calc}
\usepackage{color,bm}
\usepackage{ulem}
\usepackage{braket}
\usepackage{hyperref}
\usepackage{graphicx,epstopdf}

  \usepackage{makecell}
\newcommand{\ehbar}{\hbar_{\text{eff}}}

\newcommand{\mtn}{\mathcal{N}}

\begin{document}

\title{Super-exponential diffusion in nonlinear non-Hermitian systems}

\author{Wen-Lei Zhao}
\email[]{wlzhao@jxust.edu.cn}
\affiliation{
School of Science, Jiangxi University of Science and Technology, Ganzhou 341000, China}
\author{Longwen Zhou}
\email[]{zhoulw13@u.nus.edu}
\affiliation{
Department of Physics, College of Information Science and Engineering, Ocean University of China, Qingdao, China 266100
}
\author{Jie Liu}
\email[]{jliu@gscaep.ac.cn}
\affiliation{
Graduate School of China Academy of Engineering Physics, Beijing 100193, China
}
\affiliation{
CAPT, HEDPS, and IFSA Collaborative Innovation Center of the Ministry of Education, Peking University, Beijing 100871, China
}
\author{Peiqing Tong}
\email[]{pqtong@njnu.edu.cn}
\affiliation{
Department of Physics and Institute of Theoretical Physics, Nanjing Normal University, Nanjing 210023, China
}
\affiliation{
Jiangsu Provincial Key Laboratory for Numerical Simulation of Large Scale Complex Systems, Nanjing Normal University, Nanjing, Jiangsu 210023, China
}
\author{Kaiqian Huang}
\affiliation{
Guangdong Provincial Key Laboratory of Quantum Engineering and Quantum Materials, SPTE,South China Normal University,Guangzhou 510006,China
}

\begin{abstract}
We investigate the quantum diffusion of a
periodically kicked particle subjecting to both nonlinearity induced self-interactions and $\mathcal{PT}$-symmetric potentials.
We find that, due to the interplay between the nonlinearity and non-Hermiticity, the expectation value of mean square of momentum scales with time in a super-exponential form $\langle p^2(t)\rangle\propto\exp[\beta\exp(\alpha t)]$, which is faster than any known rates of quantum diffusion. In the $\mathcal{PT}$-symmetry-breaking phase, the intensity of a state increases exponentially with time, leading to the exponential growth of the interaction strength. The feedback of the intensity-dependent nonlinearity further turns the interaction energy into the kinetic energy, resulting in a super-exponential growth of the mean energy. These theoretical predictions are in good agreement with numerical simulations in a $\cal{PT}$-symmetric nonlinear kicked particle. Our discovery establishes a new mechanism of diffusion in interacting and dissipative quantum systems. Important implications and possible experimental observations are discussed.
\end{abstract}
\date{\today}


%
%
%

\maketitle

\begin{figure}[htbp]
\begin{center}
\includegraphics[width=8.5cm]{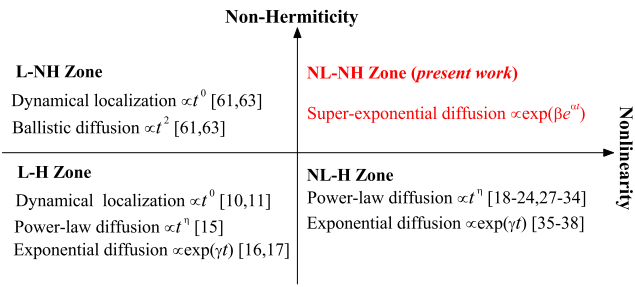}
\caption{Schematic diagram for quantum diffusion in different situations. Left half-plane: linear (L) zone; Right half-plane: nonlinear (NL) zone; Lower half-plane: Hermitian (H) zone;
Upper half-plane: non-Hermitian (NH) zone.
\label{schematic}}
\end{center}
\end{figure}

{\color{blue}\textit{Introduction.---}}
Diffusion of particles is of fundamental importance in many disciplines of physics, e.g., statistical physics and condensed matter physics. Its mechanism governs the conductivity of electrons~\cite{Anderson58}, the spin transport~\cite{Stejskal65}, the heat transfer~\cite{Kittel80}, and the Fermi acceleration of cosmic ray particles~\cite{Fermi49,Leonel10,Karlis06,Lemoine19,Jose86}, just to name a few.
In classical domain, a seminal result of the random motion of Brownian particles is normal diffusion~\cite{Einstein05}, which is characterized by the linear growth ($\propto t$) of the second moment of observable.
Quantum mechanically, the random diffusion of microscopic particles in disordered potential is totally suppressed by quantum interference, leading to the well-known Anderson localization (AL)~\cite{Anderson58}. Its analog in momentum space is the dynamical localization (DL) (see the L-H zone in Fig.~\ref{schematic}), which occurs in chaotic systems periodically driven by impulsive fields~\cite{Casati79,Fishman82,Izrailev90,Frahm97,Tian11}.

In Hermitian case, periodically-driven systems exhibit interesting diffusion behaviors, such as power-law diffusion $\propto t^{\eta}$~\cite{Fang16} and exponential diffusion $\propto e^{\gamma t}$~\cite{JWang11,Hailong13} (see the L-H zone in Fig.~\ref{schematic}), which are originated from the quantum resonance.
In the past two decades, extensive studies have been concentrated on the diffusion process in complex systems, where  the disorder and nonlinearity may coexist~\cite{Molina,
Shepel93,Pikovsky,Ignacio09,Flach08,Flach09a,Flach10b}. Nonlinear effects appear in a broad range of systems, for instance the Bose-Einstein condensates~\cite{Pitaevskii}  and the nonlinear optics~\cite{Siegman}.
Up to now, a wide spectrum of diffusion processes, from power-law diffusion~\cite{Fishman12,Flach13,Veksler09,Pikovsky11,Michaely12,Vakulchyk18,Cherroret14,Gligoric13} to exponential diffusion~\cite{Mieck05,Guarneri17,Zhao16,Zhao19JPA} have been found in nonlinear systems (see the NL-H zone in Fig.~\ref{schematic}).

A common condition of for the appearance of these diffusion processes is the assumption of Hermiticity of quantum mechanics.
Even without Hermiticity, a new class of system with $\cal{PT}$ symmetry possesses the real eigenvalues as well~\cite{Moiseyev2011,Bender1998,Bender2002}.
The non-Hermitian Hamiltonian can be used to describe non-equilibrium relaxation problems~\cite{Verbaarschot85}, optical transport in lossy media~\cite{Xiao17,Freilikher94}, and open quantum systems~\cite{Bender2007}. Thus it has been a subject of extensive theoretical~\cite{Cao2015,Ashida20,Gong19,Gong13,Makris08,Longhi09a,Longhi09b} and experimental studies~\cite{Ganainy2018,Keller97,Xiao16,Li16,Kreibich16,Kreibich13,Kreibich14,Hang18}. The non-Hermitian extension of Floquet-driven systems stimulates fruitful studies on the quantum diffusion behavior, where DL phenomenon and ballistic diffusion have been reported~\cite{Longhi17} (see the L-NH zone in Fig.~\ref{schematic}).
In this context, the quantum diffusion in a system with non-Hermiticity and nonlinearity deserves urgent investigations~\cite{Suchkov16,Zhao19,Musslimani08,Konotop,Midya17,Komissarova19,Ziheng19,Tsoy18}.

In this work, we investigate the wavepacket dynamics in a Floquet system, where both the $\mathcal{PT}$-symmetry and nonlinearity are periodically modulated in time. In the broken $\mathcal{PT}$-symmetry phase, for which the quasi-energies become complex, the wavepackets diffuse in a way that the mean square of momentum is the exponent of exponent of time, i.e., $\langle p^2(t)\rangle\propto\exp[\beta e^{\alpha t}]$. To the best of our knowledge, this is the first report of a super-exponential diffusion (SED). The underlying physics is due to the coexistence of two facts: i) the exponentially-fast growth of the intensity of wave function due to the non-Hermiticity; ii) the positive feedback mechanism of the intensity-dependent nonlinearity. Our theoretical prediction of the law of the SED is consistent with numerical results.

{\color{blue}\textit{Model.---}}The system we consider is a quantum kicked particle in an infinite square well~\cite{Hu99,Liujie00}. We make an extension to the $\cal{PT}$-symmetric kicking potential which, in contrast to Hermitian kicking, induces exotic transport behaviors~\cite{Zhao19,Zhao20}. The interatomic interaction is described by a mean-field nonlinear term, which is temporally modulated by delta kicks. The Hamiltonian in dimensionless units reads
\begin{equation}\label{KParticle}
{\rm H}  = \frac{{p}^2}{2}+V(x)+ V_{\rm K}(x) \sum_n\delta(t-n) + {\rm H}_{\rm I}  \sum_n\delta(t-n)\;,
\end{equation}
where
\begin{equation}\label{well}
V(x)=
\begin{cases}
0 & \text{if } |x|< \frac{L}{2},\\
+\infty & \text{otherwise},
\end{cases}
\end{equation}
\begin{equation}\label{Kickpotential}
V_{\rm K}(x)= K\left[ \cos(x) + i\lambda\sin(x)\right]\;,
\end{equation}
and
\begin{equation}\label{Kickpotential}
{\rm H}_{\rm I}= g_{0}|\psi(x,t)|^2\;.
\end{equation}
Here $p = -i \ehbar \partial /\partial x$ is the momentum operator, $x$ is the coordinate, and $\ehbar$ denotes the effective Planck constant with the commutation relation $[x,p]=i \ehbar$.  The parameter $L$ controls the width of the infinite square well. In the kicking potential $V_{\rm K}(x)$, the parameter $K$ indicates the strength of its real part, and $\lambda$ is the strength of its imaginary part. The parameter $g_0$ controls the nonlinear interaction strength. It is worth noting that the delta-kick nonlinearity induces rich physics, such as exponential instability~\cite{Mieck05,Guarneri17,Zhao16,Zhao19JPA} and dispersionless dynamics of wavepackets~\cite{Goussev}, which are absent in systems with static nonlinear interactions~\cite{Zhang04}. In addition, this kind of system with rigid boundary condition has served as a prototype for investigating the Fermi acceleration of particles in a
a cosmic ray in astrophysical plasmas~\cite{Fermi49,Leonel10,Karlis06,Lemoine19,Jose86}.
Therefore, the $\mathcal{PT}$-symmetric extension of this system is of broad interest.

Let $|\varphi_n\rangle$ be the eigenstate of the unperturbed Hamiltonian ${\rm H}_0 = p^2/2 + V(x)$, with ${\rm H}_0 |\varphi_n\rangle = E_n|\varphi_n\rangle$. In the representation of $| \varphi_n \rangle $, an  arbitrary state can be expressed as $|\psi (t)\rangle = \sum_{n=0}^{+\infty} \psi_n(t)| \varphi_n \rangle$, with
$\psi_n(t)$ being the component of the eigenstate $|\varphi_n \rangle$. The initial state is taken as the ground state, i.e., $\psi(x,0)=\sqrt{2 / L} \cos (\pi x / L)$.
The time evolution of quantum state over a period is governed by $|\psi(t+1)\rangle = U |\psi (t)\rangle$.
Due to the periodic kicking, the Floquet operator has the expression $U= U_{\rm f}U_{\rm K}$
where  the kicking evolution operator is $U_{\rm K}=\exp\left[ -i V_{\rm K}(x)/\ehbar -  i{\rm H}_{\rm I}(x,t)/\ehbar  \right]$,
and the free evolution operator is $U_{\rm f}= \exp(-i{p}^2/{2\ehbar})$. Without loss of genelarity, we consider the case with $L=2\pi$, for which the particle can experience the kicking potential of a full period of $2\pi$.

{\color{blue}\textit{Numerical simulations.---}}A commonly used quantity to characterize the quantum diffusion in momentum space is the
expectation value of kinetic energy
\begin{equation}\label{Energy}
\langle p^2(t) \rangle = {\sum_n p_n^2 |\psi_n(t)|^2}/{\mtn}\;,
\end{equation}
where  the norm of the quantum state is $\mtn (t)=\sum_n |\psi_n(t)|^2$. This quantity coincides with the expectation value of kinetic energy up to a factor of $1/2$.
Note that in the phase that breaks the $\mathcal{PT}$-symmetry, the norm $\mtn (t)$ could increase exponentially with time.
Thus, the above definition of expectation value drops the contribution of the norm $\mtn (t)$.

\begin{figure}[t]
\begin{center}
\includegraphics[width=8.5cm]{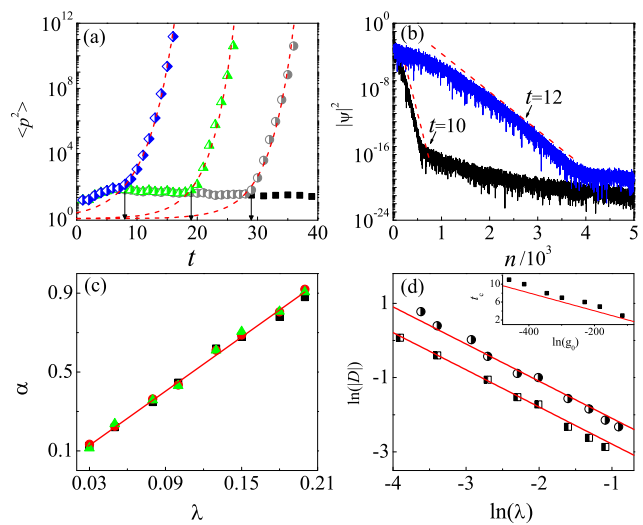}
\caption{(a) Time dependence of the mean energy $\langle p^2\rangle$  for $\lambda=0.05$ with $g_0 =0$ (squares), $10^{-7}$ (circles), $10^{-4}$ (triangles), and $0.1$ (diamonds).
Dashed lines (in red) denote the fitting function of the form $\langle p^2\rangle \approx \exp \left( \beta e^{\alpha t}\right)$. Arrows mark the threshold time $t_c$ for the appearance of the super-exponential diffusion.
(b) Probability density distribution in eigenstate space with $\lambda=0.05$ and $g_0=0.1$ at times $t=10$ (black curve) and 12 (blue curve). Dashed lines (in red) indicate the exponential profile of the form $|\psi_n|^2 \propto \exp(-n/\xi) $ with $\xi$ being the localization length. (c) The growth rate $\alpha$ of the mean energy versus $\lambda$ for $g_0=0.1$ (squares), $10^{-3}$ (circles) and $10^{-5}$ (triangles). Red line indicates our theoretical prediction $\alpha \propto K\lambda/\ehbar$ in Eq.~\eqref{EEFunct}. Other parameters are $K=5$,  $\ehbar=0.5$, and $L=2\pi$. (d) The growth rate $|D|$ versus $\lambda$ with $K=5$ for $\ehbar =0.5$ (circles) and 0.25 (squares). Red lines indicate the theoretical prediction $|D|= \ehbar/(2K \lambda)$ in Eq.~\eqref{Criticaltime}. Inset: the threshold time $t_c$ versus $\ln(g_0)$ for $K=5$, $\lambda = 0.5$ and $\ehbar=0.1$. Solid line indicates our theoretical prediction in Eq.\eqref{Criticaltime}.\label{diffusion}}
\end{center}
\end{figure}

In the present work, we investigate both numerically and theoretically the time dependence of the mean kinetic energy  $\langle p^2(t) \rangle$.
We consider the case that the system is in the $\mathcal{PT}$-symmetry-breaking phase, which is guaranteed by setting the value of the imaginary part of kicking potential to be sufficiently large.
Figure.~\ref{diffusion}(a) shows that, for $g_0=0$, the mean kinetic energy is suppressed during the time evolution, which is just the phenomenon of DL. Interestingly, for a specific value of the nonlinear strength (e.g., $g_0=0.1$), the mean kinetic energy follows that of the linear case $g_0=0$ for time smaller than a threshold value $t_c$, beyond which it increases in a super-exponential way.
Such an intrinsic phenomenon occurs even if the nonlinear strength is very small e.g., $g_0=10^{-7}$.
More importantly, we theoretically find the law of the super-exponential growth of mean kinetic energy
\begin{equation}\label{Supeexp}
\langle p^2(t) \rangle \approx
\exp\left( \beta e^{\alpha t}\right)\;,
\end{equation}
with $\alpha \propto K \lambda/\ehbar$ and $\beta \propto g_0^2$. As a further step, we numerically calculate the growth rate $\alpha$ for different $\lambda$, as shown in Fig.~\ref{diffusion}(c). One can see that the growth rate $\alpha$ increases linearly with $\lambda$. Moreover, its change with respect to $g_0$ is negligible, coinciding with
our theoretical prediction that $\alpha \propto K \lambda/\ehbar$.
The corresponding probability distribution in eigenstate space is shown in Fig.~\ref{diffusion}(b), which demonstrates the exponentially decayed profile, i.e., $|\psi_n|^2 \propto \exp(-n/\xi)$ with $\xi$ being the localization length. Taking into account $\langle p^2(t)\rangle \propto \ehbar^2 \xi^2$, the localization length will increase in the super-exponential way, which is dramatically different from the phenomenon of the DL.

From Fig.~\ref{diffusion}(a), one can also see that the threshold time $t_c$ for the appearance of the SED decreases with the increase of nonlinear strength $g_0$.
Numerical results of $t_c$ for different $g_0$ are depicted in the inset of Fig.~\ref{diffusion}(d), which demonstrates the good agreement with the analytical formula in Eq.~\eqref{Criticaltime}.
To further confirm our analytical analysis, we numerically investigate the growth rate $D$ of $t_c$ for different $\lambda$. The numerical results are in good agreement with the theoretical prediction, i.e., $D =- \ehbar/(2K\lambda)$ [see Fig.~\ref{diffusion}(d)].
We want to stress that we have also numerically investigated the system with the periodic boundary condition, which is just the $\cal{PT}$-symmetric extension of the kicked rotor model.  This system exhibits the same SED, which can be predicted by our theory as well. Thus, the SED is expected to be a universal phenomenon induced by the coexistence of non-Hermiticity and nonlinearity.

{\color{blue}\textit{Theoretical analysis.---}} We concentrate on the case of $\mathcal{PT}$-symmetry-breaking phase, i.e., $K\lambda/\ehbar \gg 1$, for which the norm exponentially increases with time $\mathcal{N} \approx \exp(K\lambda t/\ehbar)$.
As an estimation, we analyze the time evolution of the quantum state at the point $x_0=\pi/2$, which corresponds to the maximal value of imaginary part of the kicking potential, i.e., $V_{\rm i} (x_0)= K\lambda \sin(x_0)=K\lambda$. After several kicking periods, the quantum state is extremely centered at $x_0$, since the action of the the imaginary kicking term of the Floquet operator $U_{\rm K}^{\rm i} (x_0)=\exp(K\lambda/\ehbar)$ on a quantum state can greatly amplify the probability amplitude of the state in $x_0$ if $K\lambda/\ehbar \gg 1$. Accordingly, the time evolution of the  probability amplitude for $x_0 = \pi/2$ is approximately given by
\begin{equation}\label{Norm}
|\psi(x_0,t)|^2 \propto \mathcal{N}(t) \propto \exp\left(2 \frac{K\lambda}{\ehbar}t\right)|\psi(x_0,0)|^2\;.
\end{equation}
As a consequence, the nonlinear interaction strength increases exponentially as,
\begin{equation}\label{NonInter}
g_0|\psi(x_0,t)|^2 \propto g_0|\psi(x_0,0)|^2  \exp\left(2 \frac{K\lambda}{\ehbar}t\right)\;.
\end{equation}

In this system, there is a competition between the non-Hermitian kicking potential and the nonlinear interaction. The non-Hermitian kicking potential leads to localization. Meanwhile, the nonlinear interaction destroys the localization. At the initial time, the nonlinear interaction at $x_0$, i.e., $g_0|\psi(x_0,0)|^2$ is much smaller than the imaginary part of the non-Hermitian kicking potential. Thus, during the short-time evolution, the dynamics of this system is governed by the non-Hermitian kicking potential.
When the nonlinear interaction strength exceeds the
imaginary kicking strength, the effects of nonlinear interaction
dominate the dynamical behavior of the system, and consequently causing the appearance of the SED. It is hence straightforward to get the threshold time $t_c$ by
\begin{equation}\label{Balance}
g_0|\psi(x_0,t)|^2 = K\lambda\;.
\end{equation}
Combining Eqs.\eqref {NonInter} and \eqref{Balance} yields the relation
\begin{equation}\label{Criticaltime}
t_c \propto -\frac{\ehbar}{2K\lambda} \ln(g_0) + \frac{\ehbar}{2K\lambda} \ln\left(\frac{K\lambda}{|\psi(x_0,0)|^2}\right)\;,
\end{equation}
which is confirmed by our numerical results~[see the inset of Fig.~\ref{diffusion}(d)].

Next, we evaluate the time dependence of the mean kinetic energy.
Previously, we developed a hybrid quantum-classical (HQC) theory to explain the exponential diffusion induced by temporally-modulated nonlinear interactions~\cite{Zhao16,Zhao19JPA}. Here, we make an extension of the theory to systems with non-Hermiticity.
Our HQC theory predicts an iterative equation of energy
\begin{equation}\label{ItrtEq}
\langle p^2(t+1)\rangle  \approx \langle p^2(t)\rangle +C g^2(t) \langle p^2(t)\rangle \;,
\end{equation}
where the time-dependent nonlinear interaction strength is defined as $g(t)= g_0 |\psi(x,t)|^2$, and $C$ is an unimportant constant.
As an estimation, we use $|\psi(x_0,t)|^2$ to replace $|\psi(x,t)|^2$, which is reasonable since  $|\psi(x_0,t)|^2$ accounts the maximal contribution.
Substituting Eq.~\eqref{NonInter} into Eq.~(\ref{ItrtEq}) yields
\begin{equation}\label{ItrtEq2}
\langle p^2(t+1)\rangle \approx \langle p^2(t)\rangle+ Cg_0^2\exp\left(\frac{4K\lambda t}{\ehbar}\right) \langle p^2(t)\rangle\;.
\end{equation}
In the continuous-time limit, the above equation yields
\begin{equation}\label{ItrtEq3}
\frac{d\ln(\langle p^2\rangle)}{dt} \approx  Cg_0^2\exp\left(\frac{4K\lambda t}{\ehbar}\right)\;.
\end{equation}
Therefore, the time dependence of the mean kinetic energy takes the form
\begin{equation}\label{EEFunct}
\langle p^2(t)\rangle \propto \exp\left[g_0^2\exp\left(\frac{4K\lambda }{\ehbar}t\right)\right]\;.
\end{equation}
The validity of our theoretical prediction is confirmed by the numerical results of the growth rate of mean energy, i.e., $\alpha \propto K\lambda/\ehbar$ [see Fig.~\ref{diffusion}(c)]. We want to stress that our study  focuses on the regime away from quantum resonance, i.e., $\ehbar\neq 4\pi r/s$ with $r$ and $s$ coprime integers. Although in the quantum main-resonance case $\ehbar= 4\pi$, the intensity of the quantum state increases in the way of Eq.~\eqref{NonInter}, the mean square of momentum does not obey the iterative equation in Eq.~\eqref{ItrtEq}. As a consequence, the quantum diffusion is not super-exponentially fast. We will leave the quantum diffusion in quantum resonance situation for further investigation.

\begin{figure}[htbp]
\begin{center}
\includegraphics[width=8.0cm]{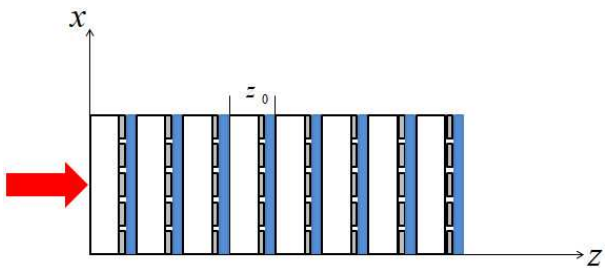}
\caption{Schematic illustration of our proposed optical system with multilayer media, where the blue layers denote Kerr media and the gray layers represent the phase gratings.\label{Optsysm}}
\end{center}
\end{figure}

{\color{blue}\textit{Experimental realization.---}}
As a further step, we propose an optical setup to emulate the
wave dynamics described by the Hamiltonian in Eq.~\eqref{KParticle}. Optical waveguides provide an ideal platform for the observation of the wavepacket transport in $\cal{PT}$-symmetric systems~\cite{Konotop,Suchkov16,Longhi17b,Song19,Feng17,Ganainy18,Zhang16,Xiao18}.
Under the paraxial approximation, the
propagation of light is governed by an equation mathematically equivalent to the Schr\"odinger equation~\cite{Sharabi18,Prange89}, where the longitude
dimension of light mimics the time variable. We consider an optical system consisting of a periodic sequence of multilayers of phase gratings and  Kerr media in the propagation direction (see Fig.~\ref{Optsysm}), which is a modification of the realization of kicked rotor model using optical settings~\cite{Fischer00,Rosen00,Agam92,Longhi17}.
It was proposed that the effect of sinusoidal and quarter-wave-shifted gratings
is described by the $\mathcal{PT}$-symmetric potential, which means that these gratings introduce the ``gain-or-loss'' mechanisms to the system~\cite{Longhi17}.
The Kerr effects of media induces a intensity-dependent nonlinear term in Eq.~\eqref{KParticle}. To realize the delta kicks in time, both the sizes of the phase grating and Kerr media in the propagation direction $z$ should be much smaller than the period of the optical sequence.
The light is trapped in transverse dimension by waveguides, which resembles the reflective boundary condition of an infinite square well in Eq.~\eqref{KParticle}. The propagation of light in such an optical system is governed the Hamiltonian in Eq.~\eqref{KParticle}.
Therefore, our finding of the SED is within reach of current experiments and may shed new light on the fundamental problems of quantum diffusion.

{\color{blue}\textit{Summary.---}}
We investigate, both numerically and analytically, the SED in a $\mathcal{PT}$-symmetric kicking system. The underlying physics of such an intrinsic phenomenon is the positive feedback mechanism of the nonlinearity, which turns the exponential-growth of intensity of wavepackets in the $\mathcal{PT}$-symmetry-breaking phase into the kinetic energy. Our theoretical prediction of the threshold time for the appearance of the SED and the law of SED are in good agreement with numerical results.
This novel behavior is of particularly importance in the field of Fermi acceleration, where the process for accelerating cosmic ray particles to large energy scales is still an open question~\cite{Leonel10,Marcowith}.

\begin{acknowledgments}
We are grateful to Jiaozi Wang for stimulating discussions.
W. Zhao. is supported by the National Natural Science Foundation of China (Grant No.12065009).
P. Tong. is supported by the National Natural Science Foundation of China (Grant No.11975126).
L. Zhou. is supported by the National Natural Science Foundation of China (Grant No.11905211), the China Postdoctoral Science Foundation (Grant No. 2019M662444), the Fundamental Research Funds for the Central Universities (Grant No. 841912009), the Young Talents Project at Ocean University of China (Grant No.861801013196), and the Applied Research Project of Postdoctoral Fellows in Qingdao (Grant No. 861905040009).
\end{acknowledgments}



\begin{thebibliography}{*}


\bibitem{Anderson58}
P. W. Anderson, Phys. Rev. {\bf 109}, 1492 (1958).


\bibitem{Stejskal65}
E. O. Stejskal, and J. E. Tanner,  Spin diffusion measurements: spin echoes
in the presence of a time-dependent field gradient. J. Chem. Phys. {\bf 42}, 288
(1965).

\bibitem{Kittel80}
C. Kittel and H. Kroemer, Thermal Physics (WH Freeman, 1980).

\bibitem{Fermi49}
E. Fermi, Phys. Rev. {\bf 75}, 1169 (1949).

\bibitem{Leonel10}
E. D. Leonel and L. A. Bunimovich, \prl {\bf 104} 224101 (2010).

\bibitem{Karlis06}
A. K. Karlis, P. K. Papachristou, F. K. Diakonos, V. Constantoudis, and P. Schmelcher, \prl, {\bf 97}, 194102 (2006).

\bibitem{Lemoine19}
M. Lemoine, \prd {\bf 99 },  083006 (2019).

\bibitem{Jose86}
J. V. Jos\'e, and R. Cordery, \prl {\bf 56}, 290 (1986).


\bibitem{Einstein05}
A. Einstein, Ann. Phys. (Leipzig), {\bf 17}  549. (1905).
English translation: Investigations on the Theory of Brownian Movement (Dover, New York) 1956.


\bibitem{Casati79}
G. Casati, B. V. Chirikov, F. M. Izrailev, and J. Ford, {\it in Stochastic
Behavior in Classical and Quantum Hamiltonian Systems},
edited by G. Casati and J. Ford, Lecture Notes in Physics
Vol. 93 (Springer, Berlin, 1979).


\bibitem{Fishman82}
S. Fishman, D. R. Grempel, and R. E. Prange, \prl {\bf49}, 509 (1982).

\bibitem{Izrailev90}
F. M. Izrailev, Phys. Rep. {\bf 196}, 299 (1990).


\bibitem{Frahm97}
K. M. Frahm and D. L. Shepelyansky, \prl {\bf 78}, 1440 (1997).


\bibitem{Tian11}
C. Tian, A. Altland, and M. Garst, \prl {\bf 107}, 074101 (2011).

\bibitem{Fang16}
P. Fang and J. Wang, Sci. China-Phys. Mech. Astron. {\bf 59}, 680011 (2016).

\bibitem{JWang11}
J. Wang, I. Guarneri, G. Casati, and J. Gong, \prl {\bf 107}, 234104 (2011).

\bibitem{Hailong13}
H. Wang, J. Wang, I. Guarneri, G. Casati, and J. Gong, \pre, {\bf 88}, 052919  (2013).

\bibitem{Molina}
M. I. Molina, \prb {\bf 58}, 12547 (1998).


\bibitem{Shepel93}
D. L. Shepelyansky, \prl {\bf70}, 1787 (1993).

\bibitem{Pikovsky}
A. S. Pikovsky and D. L. Shepelyansky, \prl
100, 094101 (2008).

\bibitem{Ignacio09}
I. Garcc\'ia-Mata and D. L. Shepelyansky, \pre {\bf 79}, 026205 (2009).

\bibitem{Flach08}
G. Kopidakis, S. Komineas, S. Flach and S. Aubry, \prl {\bf 100}, 084103 (2008).


\bibitem{Flach09a}
S. Flach, D. O. Krimer, and Ch. Skokos, \prl {\bf 102}, 024101 (2009).

\bibitem{Flach10b}
Ch. Skokos and S. Flach, \pre {\bf 82}, 016208 (2010).

\bibitem{Pitaevskii}
L. P. Pitaevskii and S. Stringari, {\it Bose-Einstein Condensation} (Oxford University Press, New York, 2003).

\bibitem{Siegman}
A. E. Siegman, {\it Lasers} (University Science Books, California, 1986).

\bibitem{Fishman12}
S. Fishman, Y. Krivolapov, and A. Soffer, Nonlinearity {\bf 25}, R53 (2012).

\bibitem{Flach13}
Ch. Skokos, I. Gkolias, and S. Flach, \prl {\bf 111}, 064101 (2013).

\bibitem{Veksler09}
H. Veksler, Y. Krivolapov and S. Fishman, \pre {\bf 80}, 037201 (2009).

\bibitem{Pikovsky11}
A. Pikovsky and S. Fishman, \pre {\bf 83}, 025201 (2011).

\bibitem{Michaely12}
E. Michaely and S. Fishman, \pre {\bf 85}, 046218 (2012).

\bibitem{Vakulchyk18}
I. Vakulchyk, M. V. Fistul, and S. Flach, \prl {\bf 122}, 040501 (2019).

\bibitem{Cherroret14}
N. Cherroret, B. Vermersch, J. C. Garreau, and D. Delande, \prl {\bf 112}, 170603 (2014).


\bibitem{Gligoric13}
G. Gli\'gori\'c, K. Rayanov, and S. Flach, EPL, {\bf 101}, 10011 (2013).


\bibitem{Mieck05}
B. Mieck and R. Graham, J. Phys. A, {\bf 38}, L139 (2005).

\bibitem{Guarneri17}
I. Guarneri, \pre {\bf 95}, 032206 (2017).


\bibitem{Zhao16}
W. Zhao, J. Gong, W. Wang, G. Casati, J. Liu, and L. Fu, \pra {\bf 94}, 053631 (2016).

\bibitem{Zhao19JPA}
W. Zhao, J. Z. Wang and W. Wang, J. Phys. A: Math. Theor. {\bf 52} 305101 (2019).



\bibitem{Moiseyev2011}
N. Moiseyev, {\it Non-Hermitian quantum mechanics}, (Cambridge University Press, Cambridge, UK, 2011).

\bibitem{Bender1998}
C. M. Bender and S. Boettcher, \prl {\bf 80}, 5243
(1998).

\bibitem{Bender2002}
C. M. Bender, D. C. Brody, and H. F. Jones, \prl {\bf 89}, 270401 (2002).

\bibitem{Verbaarschot85}
J. J. M. Verbaarschot, H. A. Weidenmuller, and M. R. Zirnbauer, Phys. Rep. 129, 367 (1985).

\bibitem{Xiao17}
L. Xiao, X. Zhan, Z. H. Bian, K. K.Wang, X. Zhang, X. P. Wang, J. Li, K. Mochizuki, D. Kim, N. Kawakami, W. Yi, H. Obuse, B. C. Sanders, and P. Xue, Nat. Phys. {\bf 13}, 1117 (2017).

\bibitem{Freilikher94}
V. Freilikher, M. Pustilnik, and I. Yurkevich, Phys. Rev. Lett. 73, 810 (1994).

\bibitem{Bender2007}
C. M. Bender, Rep. Prog. Phys. {\bf 70}, 947 (2007).

\bibitem{Cao2015}
H. Cao and J. Wiersig, \rmp {\bf 87}, 61 (2015).

\bibitem{Ashida20}
Y. Ashida,  Z. Gong,  and M. Ueda, arxiv:2006.01837 (2020).

\bibitem{Gong19}
D. Zhang, Q. Wang, and J. Gong, \pra {\bf 99}, 042104 (2019).

\bibitem{Gong13}
J. Gong and Q. Wang, J. Phys. A, {\bf 46}, 485302 (2013).



\bibitem{Makris08}
K. G. Makris, R. El-Ganainy, D. N. Christodoulides, and Z. H. Musslimani, \prl {\bf 100}, 103904 (2008).

\bibitem{Longhi09a}
S. Longhi,  \prl {\bf 103}, 123601 (2009).

\bibitem{Longhi09b}
S. Longhi, \prb {\bf 80}, 235102 (2009).

\bibitem{Ganainy2018}
R. El-Ganainy, K. G. Makris, M. Khajavikhan, Z. H. Musslimani, S. Rotter, and D. N. Christodoulides, Nat. Phys. {\bf 14}, 11 (2018).


\bibitem{Keller97}
C. Keller, M. K. Oberthaler, R. Abfalterer, S. Bernet, J. Schmiedmayer, and A. Zeilinger, \prl {\bf 79}, 3327 (1997).

\bibitem{Xiao16}
Z. Zhang, Y. Zhang, J. Sheng, L. Yang, M. Miri, D. N. Christodoulides, B. He, Y. Zhang, and M. Xiao, \prl {\bf 117}, 123601 (2016).

\bibitem{Li16}
J. Li, A. K. Harter, J. Liu, L. d. Melo, Y. N. Joglekar, L. Luo, Nat. Commun. {\bf 10},  855 (2019).

\bibitem{Kreibich16}
M. Kreibich, J. Main, H. Cartarius, and G. Wunner, \pra  {\bf 93}, 023624 (2016), and references therein.

\bibitem{Kreibich13}
M. Kreibich, J. Main, H. Cartarius, and G. Wunner, \pra {\bf 87}, 051601(R) (2013).

\bibitem{Kreibich14}
M. Kreibich, J. Main, H. Cartarius, and G. Wunner, \pra {\bf 90}, 033630 (2014).

\bibitem{Hang18}
C. Hang, and G. Huang, \pra {\bf 98}, 043840  (2018).



\bibitem{Longhi17}
S. Longhi, \pra {\bf 95}, 012125 (2017).

\bibitem{Suchkov16}
S. V. Suchkov, A. A. Sukhorukov, J. H. Huang, S. V. Dmitriev, C. H. Lee, and Y. S. Kivshar, Laser Photonics Rev. {\bf 10}, 177 (2016).


\bibitem{Zhao19}
W. Zhao,  J. Wang, X. Wang, and P. Tong, \pre {\bf 99}, 042201 (2019).


\bibitem{Musslimani08}
Z. H. Musslimani, K. G. Makris, R. El-Ganainy, and D. N. Christodoulides, \prl {\bf 100}, 030402 (2008).

\bibitem{Konotop}
V. V. Konotop, J. Yang, and D. A. Zezyulin, \rmp {\bf 88},
035002 (2016).

\bibitem{Midya17}
B. Midya and V. V. Konotop, \prl {\bf 119}, 033905 (2017).

\bibitem{Komissarova19}
M. V. Komissarova, V. F. Marchenko, and P. Yu. Shestakov, \pre {\bf 99}, 042205 (2019).

\bibitem{Ziheng19}
Z. Zhou and Z. Yu, \pra {\bf 99}, 043412  (2019).

\bibitem{Tsoy18}
E. N. Tsoy, F. Kh. Abdullaev, and V. E. Eshniyazov, \pra
{\bf 98}, 043854 (2018).

\bibitem{Hu99}
B. Hu, B. Li, J. Liu, and Y. Gu, \prl {\bf 82} 4224 (1999).

\bibitem{Liujie00}
J. Liu, T. Cheng, and S. Chen, Commun. Theor. Phys. (Beijing, China) {\bf 33}, 15 (2000).

\bibitem{Zhao20}
K. Q. Huang, J. Z Wang, W. L. Zhao, and J. Liu, submitted to J. Phys.: Condens. Matter.

\bibitem{Goussev}
A. Goussev, P. Reck, F. Moser, A. Moro, C. Gorini, and K Richter, \pra {\bf 98}, 013620 (2018).

\bibitem{Zhang04}
C. Zhang, J. Liu, M. G. Raizen, and Q. Niu, \prl {\bf 92}, 054101 (2004).



\bibitem{Longhi17b}
S. Longhi,  EPL {\bf 120}, 64001 (2017).

\bibitem{Song19}
W. Song, W. Sun, C. Chen, Q. Song, S. Xiao, S. Zhu, and T. Li, \prl {\bf 123}, 165701 (2019).

\bibitem{Feng17}
L. Feng, R. El-Ganainy, and L. Ge, Nat. Photonics {\bf 11}, 752 (2017).

\bibitem{Ganainy18}
R. El-Ganainy, K. G. Makris, M. Khajavikhan, Z. H. Musslimani, S. Rotter, and D. N. Christodoulides, Nat. Phys. {\bf 14}, 11  (2018).

\bibitem{Zhang16}
Z. Zhang, Y. Zhang, J. Sheng, L. Yang, M-Ali Miri, D. N. Christodoulides, B. He, Y. Zhang, and M. Xiao, \prl {\bf 117}, 123601 (2016).

\bibitem{Xiao18}
Z. Zhang, D. Ma, J. Sheng, Y. Zhang, Y. Zhang, and M. Xiao,  J. Phys. B: At. Mol. Opt. Phys. {\bf 51}  072001 (2018).


\bibitem{Sharabi18}
Y. Sharabi, H. Sheinfux, Y. Sagi, G. Eisenstein, and M. Segev, \prl {\bf 121} 233901 (2018).

\bibitem{Prange89}
R. E. Prange and S. Fishman, \prl {\bf 63}, 704 (1989).

\bibitem{Fischer00}
B. Fischer, A. Rosen, A. Bekker, and S. Fishman, \pre {\bf 61}, R4694(R) (2000).

\bibitem{Rosen00}
A. Rosen, B. Fischer, A. Bekker, and S. Fishman, J. Opt. Soc. Am. B {\bf 17}, 1579 (2000).

\bibitem{Agam92}
O. Agam, S. Fishman and R.E. Prange,  \pra {\bf 45}, 6773 (1992).

\bibitem{Marcowith}
A. Marcowith, A. Bret, A. Bykov, M. Dieckman, L. Drury, B. Lemb\`ege, M. Lemoine, G. Morlino, G. Murphy, G. Pelletier, I. Plotnikov,
B. Reville, M. Riquelme, L. Sironi, and A. Stockem Novo, Rep. Prog. Phys. {\bf 79}, 046901 (2016).

\end{thebibliography}
\end{document}